# High Resolution Multispectral Spatial Light Modulators based on Tunable Fabry-Perot Nanocavities

*Shampy Mansha[⊥], Parikshit Moitra[⊥], Xuewu Xu[⊥], Tobias W. W. Mass[⊥], Rasna Maruthiyodan Veetil , Xinan Liang, Shi-Qiang Li, Ramón Paniagua-Domínguez\* and Arseniy I. Kuznetsov\**

*Institute of Materials Research and Engineering, A\*STAR (Agency for Science, Technology and Research), 138634, Singapore*

\*Email: ramon_paniagua@imre.a-star.edu.sg

\*Email: arseniy_kuznetsov@imre.a-star.edu.sg

**Abstract**: Spatial light modulators (SLMs) are the most relevant technology for dynamic wavefront manipulation. They find diverse applications ranging from novel displays to optical and quantum communications. Among commercial SLMs for phase modulation, Liquid Crystal on Silicon (LCoS) offers the smallest pixel size and, thus, the most precise phase mapping and largest field of view (FOV). Further pixel miniaturization, however, is not possible in these devices due to inter-pixel cross-talks, which follow from the high driving voltages needed to modulate the thick liquid crystal (LC) cells that are necessary for full phase control. Newly introduced metasurface-based SLMs provide means for pixel miniaturization by modulating the phase via resonance tuning. These devices, however, are intrinsically monochromatic, limiting their use in applications requiring multi-wavelength operation. Here, we introduce a novel design allowing small pixel and multi-spectral operation. Based on LC-tunable Fabry-Perot nanocavities engineered to support multiple resonances across the visible range (including red, green and blue wavelengths), our design provides continuous $2\pi$ phase modulation with high reflectance at each of the operating wavelengths. Experimentally, we realize a device with 96 pixels (~1μm pitch) that can be individually addressed by electrical biases. Using it, we first demonstrate multi-spectral programmable beam steering with FOV~18º and absolute efficiencies exceeding 40%. Then, we reprogram the device to achieve multi-spectral lensing with tunable focal distance and efficiencies ~27%. Our design paves the way towards a new class of SLM for future applications in displays, optical computing and beyond.

**Introduction**

Spatial light modulators (SLMs)[1–4] are devices that can spatially modulate the amplitude, phase or polarization of light, which makes them crucial components in a wide range of applications, from emerging holographic and AR/VR displays[5] to bio-imaging, free-space optical communications or quantum computing, to name a few[3]. Among SLMs, phase-only devices typically employ liquid crystals[6] (LCs) to control the phase retardation of light on individual pixels. Due to their anisotropic molecular morphology, collectively aligned LC molecules can create a birefringent medium with an anisotropic refractive index. In the case of a nematic LC, this anisotropy can reach values as large as $\Delta n = 0.2$-$0.4$ in the visible spectral range (here, $\Delta n = n_e - n_o$ is the difference between the extraordinary and ordinary refractive indices, $n_e$ and $n_o$, respectively). Importantly, the alignment of the LC molecules can be reoriented towards a particular direction by applying an external electric field and consequently the anisotropic refractive index can be dynamically controlled. By biasing individual pixels, the LC orientation can be locally modified, creating a refractive index landscape that translates into different local phase retardations experienced by an impinging light field. This simple, yet powerful approach imposes a certain thickness for the LC layer to be able to span the full $2\pi$ phase range. This turns into high driving voltages needed to rotate the LC molecules and ultimately limits the ability to reduce the pixel size without incurring significant cross talks that render the local phase unpredictable[6]. Currently the smallest pixel size commercially available for LCoS[1,2,7] devices is more than 3 μm, a figure that has been stagnant for almost the last decade. Large pixel sizes not only result in narrower field of view (FOV), but also limit the ability to properly map sharp variations of phase in a given profile, which is of critical importance in certain applications, such as quantum optics or microscopy.

Recently, a novel approach to realize SLM devices has emerged by using actively tunable metasurfaces[8,9]. These are planar arrays of nanostructures where the phase retardation comes from their resonant interaction with light[10-21]. Active tuning of amplitude, phase or polarization is achieved either by modifying the optical properties of these meta-atoms or those of their surroundings[22,23]. Recent studies have demonstrated the use of phase change materials[24–30], transparent conducting oxides[8,31–33], two-dimensional materials[34–38], semiconductors[39–41] and liquid crystals[9,42–46] as tuning media, with very few examples achieving individual pixel control[8,9,33]. Because of their large birefringence, mature fabrication technology and negligible losses in the visible range, LC are particularly interesting when aiming for display applications. As the phase retardation comes from the metasurface itself and the LC acts only as a tunable environment, its thickness can be significantly reduced, in turn helping to shrink the pixel size[9].

So far, however, these metasurface-based devices are monochromatic in nature and their extension to multi-spectral operation poses significant design challenges that are yet to be addressed.

Here, we introduce a novel SLM that employs LC-tunable Fabry-Perot (FP) nanocavities as pixels, to simultaneously address the limitations of conventional LC-SLMs, as well as those of metasurface-based ones. Previously, this type of structure, in the non-miniaturized version, has been used to realize tunable filters[47–50]. The device, in particular, (i) consists of a thin (sub-micron) LC cell, as to improve the response time and reduce the inter pixel cross talk, (ii) comprises small (~1 μm) pixels to achieve wide FOV (18º) and (iii) can operate simultaneously at multiple wavelengths, including red (R), green (G) and blue (B) wavelengths in the visible spectrum. This device, referred to as Fabry-Perot spatial light modulator (FP-SLM), enables high reflectance and large phase modulations of $2\pi$ at these multiple wavelengths simultaneously. As shown next, we experimentally realize the device and configure it to achieve efficient programmable wide-angle beam steering and tunable-focus lensing.

**Device Design**

Figure 1(a), shows a schematic of our FP-SLM device, which consists of an array of FP nanocavities, formed by sandwiching a thin layer of LC between two partially reflective mirrors. The device comprises two sets of conducting electrodes: a thin layer (23nm) of indium-tin-oxide (ITO) on top, serving as a common electrode and a set of linear pixelated electrodes formed by patterning a 150nm thick layer of Aluminum (Al) at the bottom (two of which are schematically shown in Fig. 1(a)). These bottom electrodes define the size of each FP nanocavity pixel and can be individually biased to electrically tune the LC orientation inside each nanocavity. The whole system is sandwiched between a glass substrate and a superstrate. Light is incident from the top,

with its electric field polarized along the x-direction, coinciding with the initial orientation of the LC director when the device is unbiased, thus experiencing the LC extraordinary refractive index ($n_e$). When the LC director is reoriented to an angle $\Theta_{LC}$ by applying an electrical bias (Figure 1 (a)), the refractive index experienced by light is modified (ending in the ordinary one, $n_0$, when the LC is fully vertically rotated, $\Theta_{LC} = 0°$), thus resulting into the modulation of the resonant frequency of the FP cavity. The LC in our device rotates in the x-z plane, determined by its initial alignment direction, and has a large birefringence $\Delta n = n_e - n_0 = 0.291$ ($n_e = 1.813$; $n_0 = 1.522$).

The two partially reflective mirrors on top and bottom that form the cavity, as shown in Figure 1(a), consist of alternating layers of high and low refractive index transparent dielectric materials TiO2/SiO2, forming distributed Bragg reflectors (DBR). The thickness ($t_{layer}$) of each dielectric layer in the DBR is optimized such that:

$$t_{layer} = \lambda/4n_{layer}, \tag{1}$$

where $n_{layer}$ is the refractive index of the layer and $\lambda$ the optimization wavelength. In particular, the thicknesses in each DBR (upper & lower) are adjusted such that the DBRs exhibit high reflectance across the whole visible wavelength range (400 nm – 800 nm). The detailed description of the optimization can be found in Methods, and the simulated reflectance in Supplementary Figure S1. After the DBRs are optimized, we study the performance of FP-SLM device shown in Figure 1(a).

For the proof of concept, we first simulate (see Methods) the device performance considering a continuous bottom Al electrode. The FP-SLM device is based on Fabry Perot nanocavities, in which the number of resonances and spectral positions across the visible wavelength range can be tuned by varying the thickness of the LC layer ($h_{LC}$). The FP-SLM reflectance spectra as a function of $h_{LC}$ are shown as a color map in Supplementary Figure S2. For our device, we choose an optimum value of $h_{LC} = 750\ nm$, which results in high reflectance resonances with large associated phase shifts at all the three RGB wavelengths. Figure 1(b) shows the simulated reflectance spectrum, which indeed displays the desired resonances at RGB wavelengths. As shown there, these resonances blue-shift when the LC is reoriented from $\Theta_{LC} = 90^0$ (in-plane, blue curve) to $\Theta_{LC} = 0^0$ (out-of-plane, purple curve). Based on this behavior, we chose the operating wavelengths of the device such that the resonance positions completely "cross" these wavelengths when the LC is switched. These are demarcated by dashed vertical lines in Figure 1(c) and are $\lambda = 460\ nm$ (for blue, B), 532 nm (for green, G) and 649 nm (for red, R), with an additional wavelength $\lambda = 600$ nm for orange (O). Importantly, LC reorientation and FP resonance shift give rise to large phase modulations at these wavelengths with high associated reflectivity. This is demonstrated in Figures 1(d)-(f), where we plot the reflectance and phase shift at the three RGB wavelengths as a function of orientation of the LC molecules, indeed showing that the high reflectance (>60%) and $\sim 2\pi$ phase shifts can be achieved for all of them.

**Experimental demonstration**

For the experimental demonstration of our FP-SLM, we fabricated a device consisting of 96 (linear) electrodes, which can be addressed individually. Each electrode has a width of $w_{Al} = 1\ \mu m$ and is separated from the neighboring pixel by a gap $G = 140\ nm$, resulting in a pixel pitch of $P = 1.14\ \mu m$ (Figure 1a). Figure 2(a) shows a camera image of the final device mounted on and wire bonded to a printed circuit board (PCB). The wires are encapsulated with a glue to protect them from any short-circuit or physical tear. The device is fabricated (see Methods) using a combination of electron beam lithography and dry etching and photolithography and wet etching for the bottom electrodes and fan-outs, respectively. Ion-assisted deposition was used to achieve the precise thickness required for the DBRs (see Supplementary Figure S1 for the experimentally measured reflectance spectra and comparison with the theoretical ones). A zoomed-in optical image on the active pixel area is shown in Figure 2(b). Figure 2(c) shows the top view scanning electron microscopy image (SEM, Hitachi SU8220) of the Al linear electrodes before the deposition of the lower DBR (the inset displaying a tilted-view, magnified SEM image of the pixels). The desired height of the LC cell is achieved using polyimide as a spacer, which is then infiltrated with the LC using capillary forces. In the final device, an LC cell thickness of $h_{LC} = 530\ nm$ was achieved, which is a bit lower than our designed LC height of 750 nm.

To characterize our device, we first corroborate that all electrodes of the fabricated device can be individually controlled by biasing each of them (square waveform input, 10V$_{rms}$ at 1 kHz) and looking at the induced intensity modulation under white light illumination in a crossed polarizer-analyzer configuration, where the LC is aligned at 45 degrees with respect to the polarizer. In this situation, the electrodes should become dark after the LC is reoriented vertically upon biasing. We find that all the electrodes are working well, as shown in Supplementary Video S1. Then, we remove the analyzer and measure the reflectance spectrum (see Methods for the details on the characterization setup) when the LC director is parallel to the polarization direction and no voltage is applied to the device. Figure 2(d) shows the obtained results and, for comparison, the simulated ones when we consider $h_{LC} = 530\ nm$ and $\theta_{LC}$ =90º. As can be seen, a good agreement between both is found, displaying three distinct resonances at three different wavelengths (viz. blue, orange and red). Note that the resonance in the green range has been shifted to orange due to offset of LC thickness in the fabricated device compared to that in the optimized model. Upon biasing all the electrodes with an increasing AC voltage V$_{rms}$ at 1kHz, the observed resonances experience the expected blue-shift, as shown in the reflectance color map in Figure 2(e). For comparison, Figure 2(f) shows the simulated results when the rotation of LC is varied from 90⁰ to 0⁰ (the expected reflectance and phase modulations at the operating wavelengths of this device with $h_{LC} = 530\ nm$ are shown in Supplementary Figure S3).

Next, we measure the phase modulation provided by this device upon biasing, using interferometric measurements (see Methods for details). Again, the same V$_{rms}$ is applied to all electrodes, in this case from 0V to 8V. The results obtained near the blue, orange and red resonant wavelengths are shown, respectively, in Figures 3(a)-(c). The corresponding simulated phase tunability as a function of the LC rotation is shown in Figures 3(d)-(f), demonstrating good agreement with the experiment. The maximum phase retardations obtained experimentally for the operating wavelengths in the blue (503nm), orange (596nm) and red (662nm) are plotted in Figures 3(g)-(i), together with the simulated results. Note that, in simulations, 640nm is taken for the red wavelength instead, which is due to the slight spectral shift and smaller tuning range achieved in the experiment for this wavelength. As seen in the figures, we experimentally achieved phase shifts close to $2\pi$ for all the three wavelengths, namely >$1.75\pi$ for blue and orange and ~$1.5\pi$ for red.

To demonstrate the capability of wavefront control at the individual pixel level, we first program the FP-SLM as a tunable-angle beam steering device. To do so, different voltage profiles are applied to the 96 electrodes so as to create 0-$2\pi$ linear phase gradients in periodically repeated super-cells. The size of these super-cells determines the diffraction angle ($\theta_D$), which, for the first (±1) orders, follows the simple formula[1] $\theta_D = \pm\sin^{-1}(\lambda/nP)$, where n is the number of pixels comprising the super-cell. When a 0-$2\pi$ linear phase profile is mapped in this super-cell, the energy channels into a single diffraction order (+1 or -1, depending on the sign of the slope of the gradient), resulting in beam steering. We first show in Figure 4(a) that, when the device is unbiased, all the reflection goes into the zeroth order (simple specular reflection in the normal direction). We then bias the electrodes with periodic voltage profiles optimized to achieve the 0-2π linear phase mapping corresponding to the -1 diffraction order in super-cells with varying dimensions (see Methods). Figures 4(b)-(d) show three particular cases, in which the super-cells comprise 5, 8 and 12 electrodes, as schematically depicted on top of the figures (see Supplementary Figure S4 for the optimized voltage profiles). The colormaps in the figure show the measured reflectance as a function of reflection angle and wavelength, in the regions near the FP-SLM resonances. As can be seen, at the selected operation wavelength the reflected intensity concentrates at a certain angle, which indeed corresponds to the -1 diffraction order. Also it can be readily seen, that reprogramming the FP-SLM with different super-cells enables tunable-angle beam steering (see Supplementary Figure S5 for tuning using a wider range of super-cells). Importantly, this is possible at all selected operating wavelengths of the device (see Supplementary Figures S6 and S7).

An important feature of the FP-SLM is the fine, continuous control over the phase and the minimized coupled amplitude, which allows to achieve high efficiency levels. This is shown in the top panels in Figures 4(e)-(g). For the 5-pixel super-cell, we measure an efficiency ~20% (defined as the absolute reflectance into the -1 order) for all three operating wavelengths. In the other two cases (i.e. 8-pixels and 12-pixels) the efficiency is higher, peaking at a maximum above 40% for the eight-pixel case when operating in the blue region. We note, moreover, the remarkable difference between reflectance into the desired (-1) diffraction order and the undesired (+1) one, as well as the fact that the beam steering angle can be switched from negative to positive angle with very similar efficiencies by reversing the linear phase gradient, as shown in the bottom panels in Figures 4(e)-(g). We also note that, for the 5-pixels super-cell operating in the red region, which corresponds to the largest steering angle (~7º), this asymmetry is reduced, which might follow from the limited phase modulation obtained for this wavelength (~1.5π) and the steep phase profile required to map the tilted wavefront. This asymmetry further reduces if 4-pixels super-cells are considered (see Supplementary Figure S8, which also presents the reflection into the zeroth order) and almost completely disappears for three-pixels (not shown here), which we take as the limit at which the device stops working. Ultimately, this translates into a maximum field of view ~18º (in the red region), corresponding to the 4-pixels case. For completeness, the simulated beam steering efficiencies are shown in Supplementary Figure S9.

Finally, to demonstrate the versatility of the FP-SLM, we re-program it to act as a tunable cylindrical lens. For that, the electrodes are configured to impart a position-dependent phase profile following the formula:

$$\phi(r) = \frac{2\pi}{\lambda}[(x^2 + f^2)^{1/2} - f], \tag{2}$$

where $x$ is the distance to the center of the lens (i.e. the center of the device) and $f$ is the lens focal distance. First, we show that by reconfiguring the device, it is possible to focus light at the same focal distance for all the three operating wavelengths (blue, orange and red). Following Eq. (2), the device needs to impart different phase profiles for each of the different wavelengths in order the keep the same $f$. Interestingly, these phase profiles can be readily seen when measuring the spatially resolved near-zone reflection spectra of the device at $z = 0$ μm ($z$ being the distance to the device), as shown in Fig. 5(a)-(c). There, the spectral position of the resonance as a function of $x$ can be easily identified and, as seen, it follows closely the shape of the phase profile of the lens, with the different Fresnel zones being apparent. Note that, indeed, the optimized profile for each of the three wavelengths is slightly different, which is needed to focus all wavelengths at a same focal distance $z = 525$ μm from the device, corresponding to a numerical aperture (NA) of NA = 0.1. Figure 5(d)-(f) show the spatially resolved reflection spectra, with the resulting focusing for the three wavelength ranges measured at the same distance of $z = 525\ \mu m$ from the device. The highest focusing intensities are measured at 499nm, 591nm and 661nm wavelengths, which are very similar to those where we achieved maximum beam steering efficiencies, as shown in Figure 4. The corresponding line profiles are presented in Fig. 5(g)-(i), displaying characteristic Airy profiles. By integrating the intensity contained within the first Airy ring and normalizing by the incident power, we estimate a total efficiency of ~21% for the blue wavelength, ~27% for the orange and ~16% for the red. These focusing efficiencies are achieved readily, without numerical optimization (see Methods). We stress again that, in order to keep the same focal distance for the different wavelengths, different phase profiles and therefore different voltages are needed. These are shown in Supplementary Figure S10 for completeness. If, instead, a single voltage profile is used, the different wavelengths focus at different distances, as shown in Supplementary Video S2. Finally, in Fig. 5(j)-(l), we show that, by reconfiguring the device, it is possible to continuously shift the focal distance to any desired value for any of the three operating wavelength ranges. We illustrate this by plotting the intensity distributions as the light propagates away from the lens when the device is reconfigured to achieve different focal distances for the orange wavelength of 591nm, namely 350 μm (NA=0.15), 500 μm (NA=0.11) and 625 μm (NA=0.09). Beyond this lensing demonstration, which seeks to underline versatility and readiness of our device, we note that it should be possible to further improve the device performance, e.g. the lens profiles, by utilizing machine learning methods, but this is however left for future work.

**Conclusions**

In summary, we have proposed an architecture to realize phase-only, multispectral reflective SLMs with reduced pixel size. The device consists of an array of FP nanocavities infiltrated with LCs, which can be individually biased to provide continuous 2π phase modulation with high associated reflectivity at RGB wavelengths. By using the phase modulation associated with the FP resonance spectral shifts, we uncouple it from the physical thickness of the LC, which can be reduced down to 530nm only. This allows us to dramatically reduce the inter-pixel cross-talk, enabling miniaturization of the pixel size down to ~1 μm. We demonstrate the concept by realizing a fully programmable device with 96 linear electrodes that can be individually addressed. First, we employ it to perform dynamic beam steering, achieving peak efficiencies >40% and a maximum FOV of ~18°. Then, we reconfigure our device to realize fixed focal distance lensing for the different operating wavelengths with efficiencies ranging from 16% to27% for NA=0.1. The same device is then used to tune the focal distance at a given wavelength, demonstrating NAs from 0.09 to 0.15. Our design could be readily extended to two dimensional SLM devices by using standard CMOS technology for integrated circuits and is amenable to other electro-optical tuning mechanisms. Moreover, we foresee that the voltage and crosstalk could be further reduced, and consequently also the pixel size, if the electrodes (or at least the top one) would be placed inside the cavity formed by the dielectric mirrors, or if one would require single wavelength operation. We believe that the proposed design methodology will have wide implications in the field of ultra-high resolution dynamic wavefront control, with numerous applications in diverse fields, ranging from solid-state LiDAR to AR/VR and holographic displays, optical communications or quantum computing, to mention some.

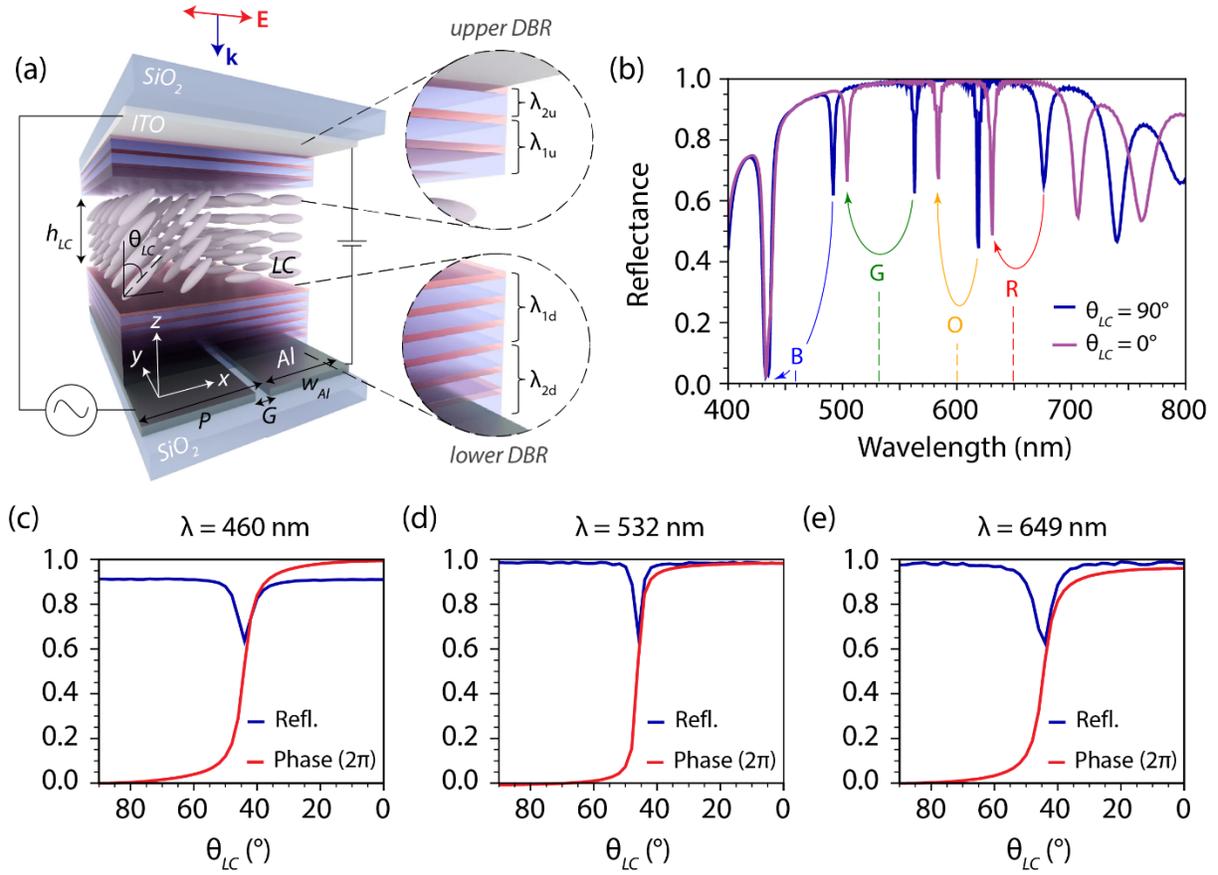

**Figure 1. Concept of FP-SLM:** (a) The schematic of the FP-SLM showing two pixels. The system comprises two partially reflecting mirrors, denoted as upper and lower DBR, a top ITO layer acting as a common electrode, a LC layer with thickness $h_{LC}$ and a set of Al electrodes forming pixels with a width $w_{Al}$, periodicity $P$ and inter-pixel gaps $G$, as specified in the main text. When bias is applied between an Al electrode and the top ITO electrode the orientation of the LC director on top of the pixel can be modified from its initial in-plane state ($\theta_{LC} = 90°$) to a certain tilted angle $\theta_{LC}$. (b) Simulated reflectance spectra for two different LC orientation angles, $\theta_{LC} = 90°$ and $0°$, for the non-pixelated FP-SLM with $h_{LC} = 750\ nm$. The dashed vertical lines represent the operating wavelengths, while the color arrows are guides to the eye indicating the initial and final positions of the resonances. (c)-(e) Simulated reflectance and phase shifts at the operating RGB wavelengths versus the LC rotation angle.

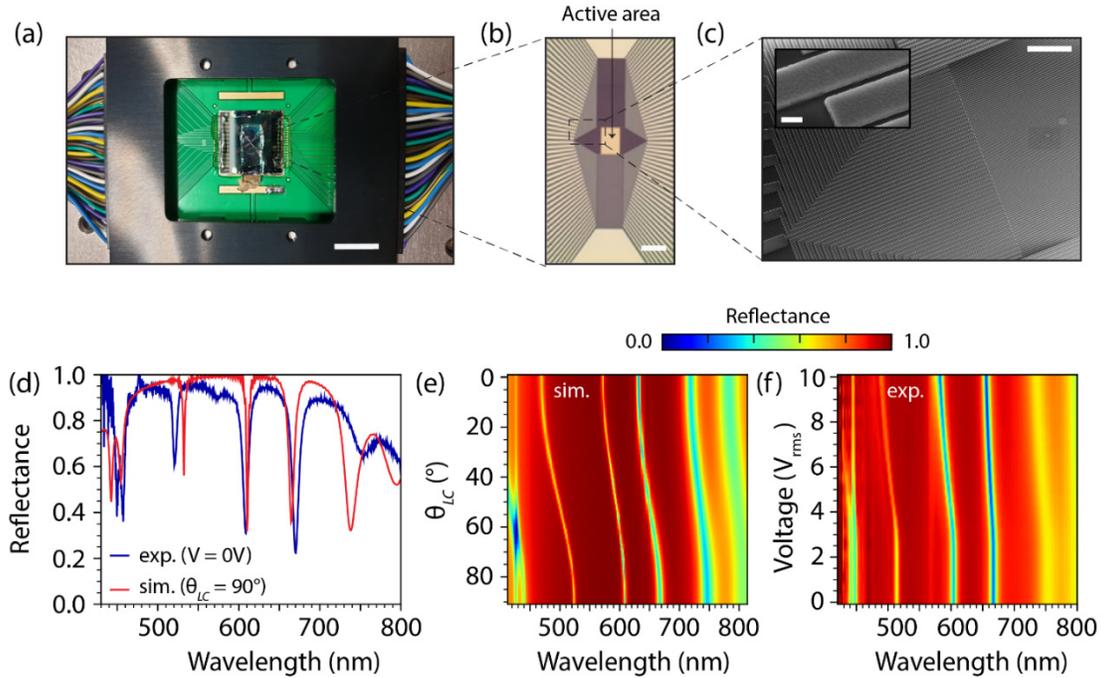

**Figure 2. Performance of the fabricated device**: (a) Optical image of the device mounted on the PCB board and connected to the voltage supply. The group of wires provides voltage to the individual electrodes (scale bar: 10 mm). (b) Optical microscope image of the fabricated 96-electrodes FP-SLM device, showing the electrodes and the active area (scale bar: 100 $\mu m$). (c) Top view SEM image of the fabricated device (scale bar: 25 $\mu m$). The inset shows a magnified tilted view SEM images (scale bar: 500 nm). The measured thickness of the Al electrodes is 158 nm. (d) Reflectance spectra for the FP-SLM. In red, simulated reflection assuming $h_{LC}$ = 530 nm and $\theta_{LC}$ = 90º. In blue, the measured spectra when no bias is applied to the device. (e) Simulated and (f) measured reflectance spectra as a function, respectively, of $\theta_{LC}$ and bias applied ($V_{rms}$). In the experiment, all pixels are simultaneously biased.

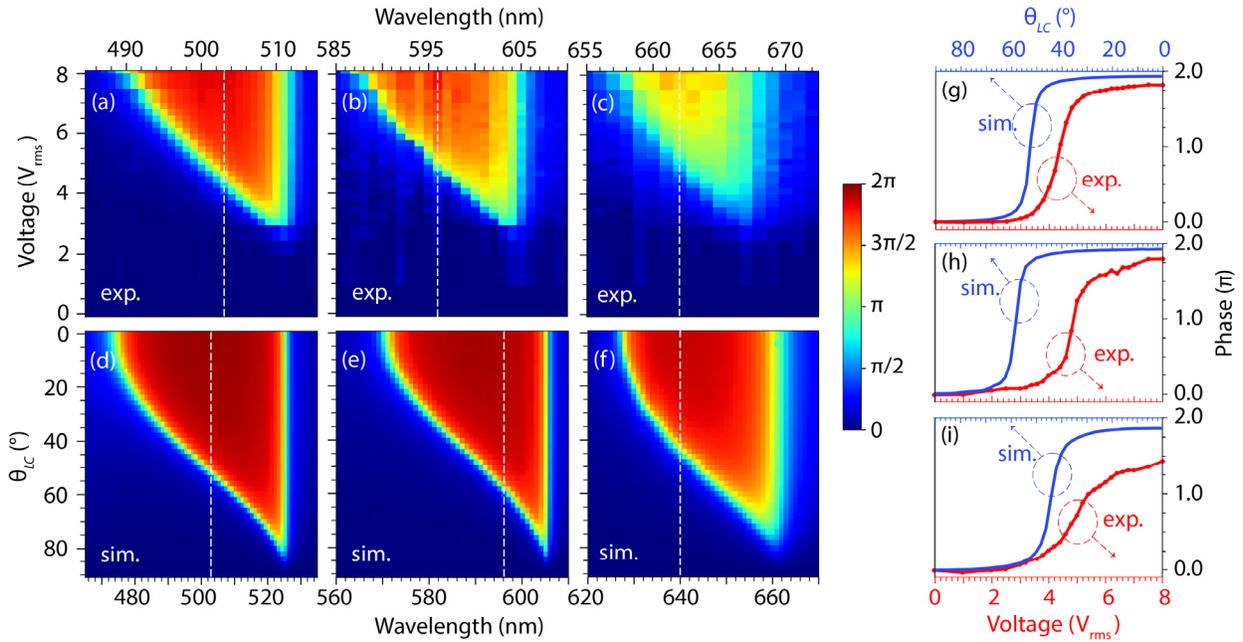

**Figure 3. Multi-spectral phase modulation of the FP-SLM** (a)-(c) Experimentally measured and (d)-(f) simulated phase shifts as a function of wavelength and, respectively, applied voltage to the FP-SLM device and LC director rotation angle ($\theta_{LC}$) around the operational wavelengths (shown as vertical dashed lines) in the blue (a, d) orange (b, e) and red (c, f) spectral regions. (g)-(i) Simulated (solid blue curve) and measured (solid red curve with symbols) at the operational wavelength in (g) the blue ($\lambda = 503\ nm$), (h) the orange ($\lambda = 596\ nm$) and (i) the red ($\lambda = 662\ nm$ in simulation, $\lambda = 640\ nm$ in experiment) spectral regions.

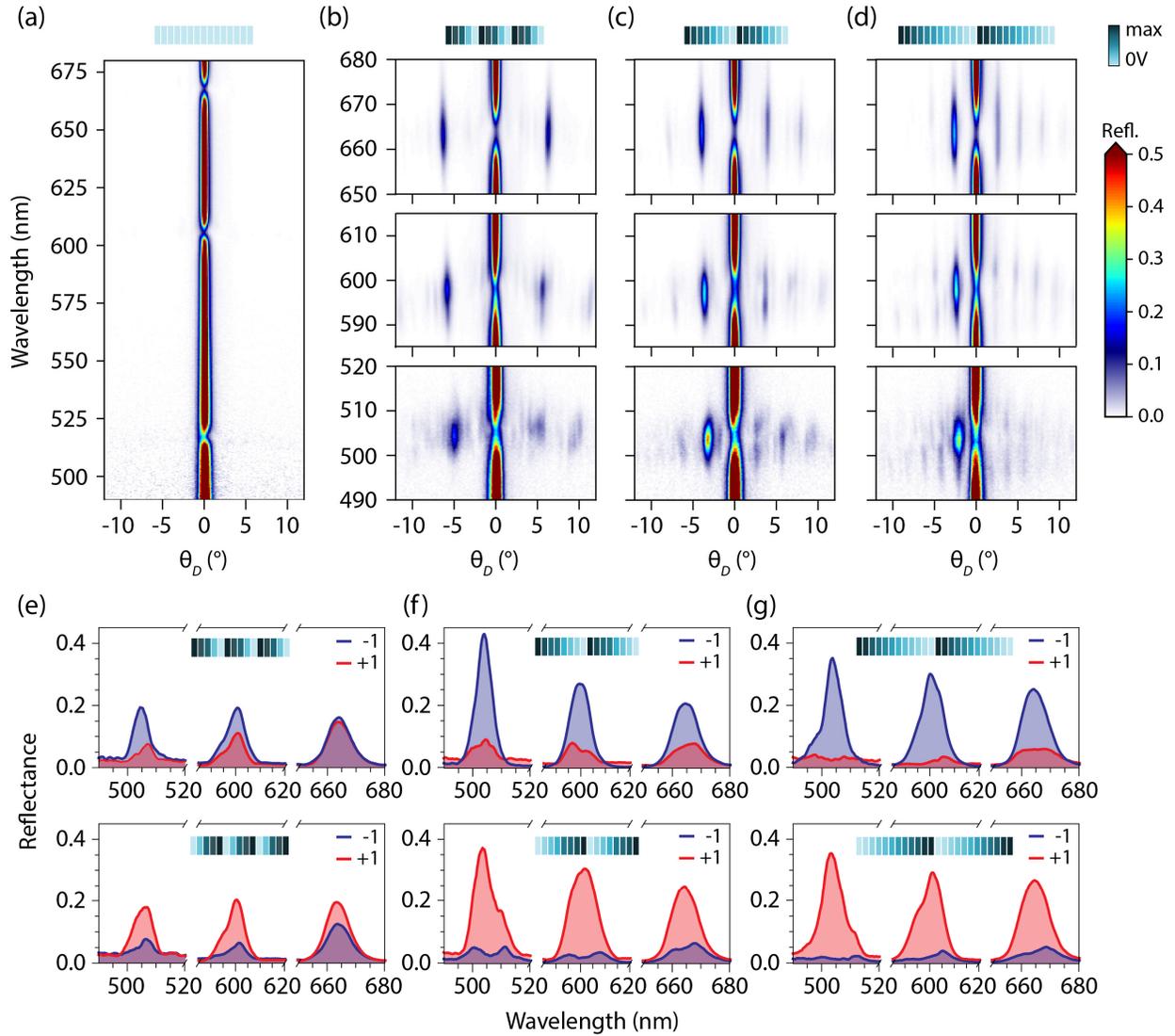

**Figure 4. Programmable beam steering with the FP-SLM**: (a)-(d) Color maps showing the measured reflectance as a function of wavelength and the reflection angle: (a) when all electrodes are unbiased, and (b)-(d) when voltage patterns are applied to the electrodes creating linear phase profiles corresponding to a blazed grating with (b) 5-pixels, (c) 8-pixels and (d) 12-pixels supercells. The voltage patterns are schematically shown on top of the panels. (e)-(g) Extracted diffraction (reflectance) efficiencies for the main orders supported by the induced gratings (-1$^{st}$ and +1$^{st}$). The top panels correspond to the cases depicted in (b)-(d), leading to power channeling into the -1$^{st}$ order, while the bottom ones correspond to the reversed angle configuration, channeling power into the +1$^{st}$ order.

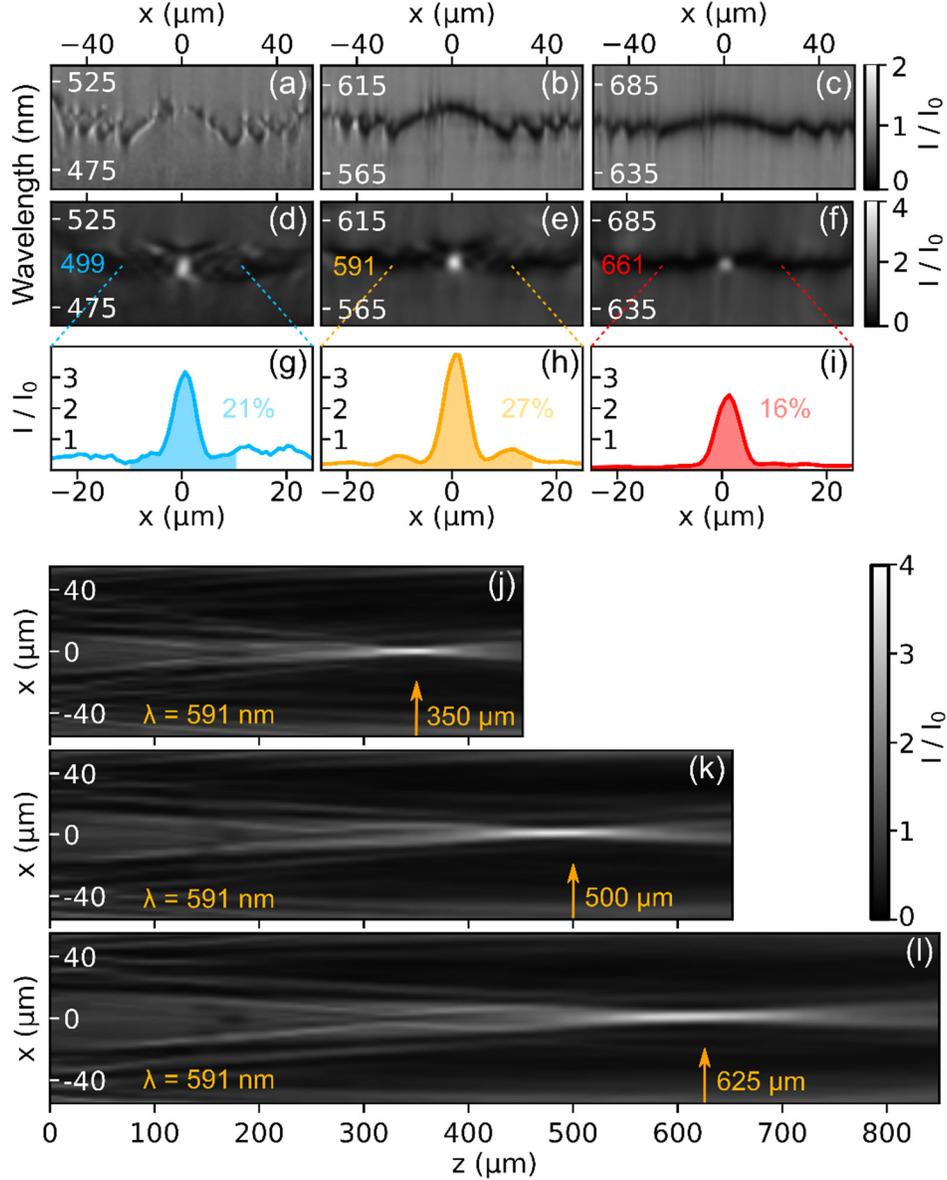

**Figure 5. Programmable lensing with the FP-SLM**: (a)-(c) Measured spatially resolved near-zone (z=0) reflectance spectra as a function of the transverse position in the device. The plots show tailored resonance positions, imparting phase profiles to achieve focusing to the same focal distance for the three operating wavelengths. (d)-(f) Measured spatially resolved reflectance spectra at z = 525μm, showing the resulting focal spots with maximum efficiencies at 499nm, 591nm and 661nm wavelengths. (g)-(i) Line profiles of the focal spots at the wavelengths indicated in (d)-(f). Focusing efficiencies corresponding to the shaded areas are 21%, 27% and 16%, respectively. (j)-(l) Measured intensity distributions as a function of the distance to the device (z) for the orange wavelength of 591 nm. Reconfigurable change of focal distance is demonstrated for three exemplary focus positions, i.e. 350 μm (j), 500 μm (k), and 625 μm (l).

**Methods**

**Simulation methodology and optimization of Distributed Bragg Reflectors**

The device design was carried out by full wave numerical simulation based on Finite Difference Time Domain method (FDTD, Lumerical Solutions). First, we calculate the reflectance of the upper & lower DBRs separately and optimize them. For that, the thicknesses of the layers in each DBR (upper & lower) is adjusted such that it can give high reflectance across the visible wavelength range (400 nm – 800 nm). To calculate the thickness of the layers we consider the refractive indice of TiO2 $n_{TiO2} = 2.48$ and that of SiO2 $n_{SiO2} = 1.46$. The upper DBR comprises the thin layer of ITO (23 nm) acting as a transparent electrode, and three pairs of TiO2/SiO2 layers, optimized according to equation (1). In particular, the first pair is optimized considering the target wavelength $\lambda_{2u} = 580\ nm$ and the next two pairs are optimized at $\lambda_{1u} = 500\ nm$. The lower DBR consists of six pairs of TiO2/SiO2, where the first three pairs are optimized at $\lambda_{1u} = 450\ nm$ and the next three at $\lambda_{2u} = 530\ nm$. The last layer in the lower DBR is a 150 nm thick Al layer acting as a bottom electrode.

For the simulation of the full FP-SLM system, Periodic Boundary Conditions (PBC) are applied in x & y directions, mimicking a single unit cell, and two PML layers are then placed on the top and bottom in the z-direction. The incident light travels through the device along z-direction with the electric field polarized along the x-direction and is then reflected back through different layers. The reflected light is then collected using a plane monitor above the topmost layer of the device. The reflected phase is calculated by averaging over the electric fields recorded at the monitor.

**Device Fabrication**

The device fabrication starts with sputtering a 150 nm thick Al on an 8 inch silicon wafer with a 1 μm thick thermal $SiO_2$ layer using physical vapor deposition system (AMAT Endura). The wafer is then diced into 20×20 mm substrates. Next, the deposited Al is patterned to create the linear electrodes and fan-outs. The electrodes in the active region are fabricated using electron beam lithography (Elionix ELS-7000) using ZEP-520A-7 positive electron beam (EB) resist, followed by reactive ion etching (RIE, Oxford OIPT Plasma Lab) of Al using chlorine ($Cl_2$) gas and boron trichloride ($BCl_3$), and EB resist as a mask. The electrode fan-outs are fabricated using Photolithography (EVG 62008 Infinity) using positive photoresist S1811, followed by wet etching of Al using the photoresist as a mask. The bottom DBR consists of six pairs of $TiO_2/SiO_2$ among which the bottom most layer, i.e. $SiO_2$, is first deposited using an ion assisted deposition tool (IAD, Oxford Optofab3000), and then followed by planarization by spin coating a 600 nm thick flowable oxide layer (FOX 24) and subsequent RIE using $C_4F_8$ gas to achieve the desired thickness (during the process, FOX 24 layer is completely etched). The rest of the 11 layers of the DBR stack are then deposited using the same deposition tool. The bottom substrate is then attached to a printed circuit board (PCB) and 96 electrodes are wire-bonded using 1 mil gold wires (where: 1 mil = 0.001 inch). A polymer adhesive is used to encapsulate the wires. The top DBR, which consists of six alternate layers of $TiO_2$ and $SiO_2$, is deposited using the same deposition tool on an ITO-glass with uniform ITO layer (23 nm). A layer of polyimide is then spin coated on the top substrate, followed by curing at 150 °C for 30 minutes on a hot plate. The cured slides are then rubbed unidirectionally with a piece of soft velvet clothes to define a preferred liquid crystal alignment direction. The rubbing strength is controlled by a commercially available rubbing machine (Holmarc Opto-Mechatronics - HO-IAD-BTR-03).

The LC cell is assembled by pressing the top (25 mm × 5 mm) against the bottom substrate (20 mm × 20 mm) using homemade press equipped with a multi-point optical profilometer. The resultant cell thickness is defined by a UV adhesive, a spacer and the pressure applied. The cell thickness is measured using a spectrometer. The nematic liquid crystal QYPDLC-001C is then encapsulated into the cell through capillary filling. The device is checked with a microscope under crossed polarizer-analyzer configuration, when the LC alignment is parallel to the polarizer to ensure the uniform extinction of LC in the gap and on the electrodes of the device.

**Electro-optical characterization**

Incident light from a halogen lamp is linearly polarized perpendicular to the 96 discrete Al electrodes. A pinhole having a diameter of 75mm (Thorlabs P75D) is used to limit the angular spread of the incident light to less than 1° from the normal. The light passes through a bright field objective lens (Nikon 20×, NA = 0.45) and is collected by the same lens after it is reflected from the device. A commercial silver mirror (Thorlabs PF10-03-P01) is used to record the incident light intensity as a reference for reflective spectrum calculation and normalization. For beam bending measurements, the back focal plane (BFP) of the objective lens is imaged to the 100μm entrance slit of a spectrograph (Andor Kymera 328i). Perpendicular to the slit, i.e. parallel to the electrodes, the light is dispersed by a blazed grating so that the back focal plane can be spectrally resolved in one direction. The image is collected by a Newton EMCCD-971 sensor array with an angular resolution of 0.16° and a spectral resolution of 0.31nm. Diffraction (reflectance) efficiencies as shown in Figure 4, are calculated by normalizing the diffracted (reflected) light from the device to the light reflected from a commercial silver mirror (after taking into account the photodetector noise effect i.e. dark current subtraction).

Similarly, in the lensing experiments the spectrally resolved real space is imaged at different distances from the device. The voltages at the 96 electrodes are set using a Texas Instruments (TI) 96-channel, 12-bit digital-to-analog converter (DAC) chip on an evaluation board (DAC60096EVM). The board is addressed by an ESP 32 microcontroller (MCU) via serial peripheral interface (SPI). The MCU, in turn, is connected to a CPU via "universal asynchronous receiver-transmitter" (UART), so that the voltage patterns can be ultimately set and optimized in a python program (see section "Device performance optimization" below for the optimization method).

**Phase retardation measurements**

The phase retardation $\varphi$ of the device is measured at different applied voltages and at different wavelengths with a Michelson interferometer using a temporal phase-shift approach and five step phase retrieval algorithm[51,52]. A supercontinuum source (SuperK EXTREME, NKT Photonics) and multi-wavelength filter (SuperK SELECT, NKT Photonic) are used as the light source. The laser beam of a given wavelength is split into two beams using a beam splitter: one beam, shone on the sample and reflected back, is referred here as the signal beam; another beam shone on a mirror mounted on a piezo-driven stage, is referred here as the reference beam. The reflected signal and the reference beams interfere and form fringes in their overlap region, which are then imaged on a camera (Thorlabs, Quantalux 2.1 MP Monochrome sCMOS Camera). The temporal phase shift is introduced by moving the mirror in the reference arm to different positions using the piezo stage, which results in creating desired optical path differences (corresponding to 0, 0.5π, π, 1.5π and 2π phase shift). The corresponding fringes are captured with the camera as $I_1$, $I_2$, $I_3$, $I_4$ and $I_5$ (where $I$ denotes the intensity) and are used to calculate the wrapped phase map according to a five-step phase retrieval algorithm as shown in equation below.

$$\varphi = arctan\left[\frac{2(I_2 - I_4)}{2I_3 - I_5 - I_1}\right]$$

For the five-step algorithm, the phase shift step between two adjacent frames should be 0.5 π, corresponding to the piezo stage moving a distance of quarter (1/4) of the wavelength. As it is a reflection geometry, the light path is doubled, therefore the moving step is one-eighth (1/8) of the wavelength. After the wrapped phase is obtained with the algorithm, it is filtered with a sin-cos filter and unwrapped to obtain the final phase map of the device. A local reference isolated from the electrodes (and hence providing a constant phase irrespective of the applied voltage), is used to calculate the relative phase-change in the area of interest.

**Device performance optimization**

*Beam steering profiles*

Properties of the fabricated devices can naturally differ slightly from the intended device design, especially in prototype development. Demonstrating the full potential of a new technology can become challenging if these differences reduce efficiency and are not compensated for. While our device is fairly close to its design, in Fig. 2 (d) and Fig. 3 in the main text, differences between simulation and experiment in reflectance and phase spectra become apparent, especially in the red wavelength region. Furthermore, as already pointed out in the main text, we noticed a slight inter-pixel crosstalk. Therefore, manual voltage setting is challenging and by doing so, one is not able to exploit and demonstrate the device performance to its full extent. Therefore, we used the python Application Programming Interface (API) of the nonlinear optimization (NLopt) toolbox[53] to optimize the voltage supercells in our bending experiments. Our setup allows a cycle time of a spectral measurement of around 0.47s. The cycle time is composed of exposure time of 0.25s, which is necessary to obtain a sufficient signal-to-noise ratio, and overhead time of 0.22s, which includes CCD readout time and other spectrometer software processing. While the cycle time constitutes a lower limit to the cost function evaluation time of our real time optimization, we add a small buffer time so that a single evaluation is still performed in less than a second. During this time, a back focal plane (BFP) spectral measurement is performed, the data are evaluated and the voltages are updated. The cost function maximizes the integral of light intensity in the target wavelength and bending angle range normalized by the intensity in the same wavelength range but across all solid angles captured by the objective. While the parameter space in an optimization of phase values is a hypersphere, the parameter space formed by 96 voltage values, having a lower bound of 0V and an upper bound of 10V, is a perfect hypercube. In order to account for rapidly varying voltages, necessary to realize phase jumps from 0 to $2\pi$ at the supercell boundaries, we employ the "Locally biased DIviding RECTangles" (DIRECT-L) algorithm, which is a global, derivative free optimization algorithm[54,55]. The algorithm works best for low aspect ratio parameter spaces, which is the case of a hypercube. It initiates all voltages at the medium value of 5V and by setting a large initial step size of 5V the whole parameter space is spun with the first divisions. Having supercells sizes ranging from 5 electrodes to 12 electrodes, optimizations require 5 to 20 minutes of time, which relates to at least 300 to 1200 function evaluations, respectively. Subsequently, in order to refine the result of the global optimization, we run a purely local optimizer using the same cost function with the result of the global optimizer as input parameter. The local derivative free algorithm used for that purpose is the "Constrained Optimization BY Linear Appoximation" (COBYLA) algorithm, also provided by the NLopt package[53,56,57]. Representative results on the obtained optimized voltages are shown in Fig. S4.

*Lensing profiles*

The aforementioned slight inter-pixel crosstalk and natural fabrication imperfections can be even a bigger challenge for more complex functionalities than beam steering. While for beam steering, optimization parameters are reduced to the supercell size, a drastic reduction of parameters is not possible for e.g. optimizing a lens profile. Symmetry would allow to half the number of electrodes, but 48 optimization parameters is still too much for a real time optimization in our experiments. Therefore, we demonstrate the readiness of our device by realizing efficient lensing without numerical optimization, purely based on knowledge about the phase profile according to formula (2) in the main text and dependence of resonance position and resulting phase from the set voltages. The voltage patterns used for the lensing experiments are provided in Fig. S10. Again, we stress that applying machine learning methods will further improve device efficiency for complex functionalities.

**Acknowledgements**

We acknowledge Son Tung Ha for his help with the optical characterization and Norhanani Jaafar for her help with SLM-PCB packaging.

**Author Contributions**

⊥S.M., P.M., X.X and T.W.W.M contributed equally to this work. R.P.-D. and A.I.K. conceived the idea. S.M. performed the numerical simulations and developed the design. P.M. conducted the nanofabrication of the device. X.X. and T.W.W.M. performed the optical characterization.

T.W.W.M. optimized the driving method and performed the device optimization. R.M.V. fabricated the liquid crystal. X.L. performed the phase measurements. S.-Q.L. helped in the fabrication of initial samples and designed the initial device driving method. R.P.-D. and A.I.K supervised the work. All authors analyzed the data and contributed to the manuscript preparation.

**Competing interests**

The authors declare no competing financial interest.

**Funding Sources**

This work was supported in part by the AME Programmatic Grant, Singapore, under Grant A18A7b0058; in part by the IET A F Harvey Engineering Research Prize 2016; and in part by the National Research Foundation of Singapore under Grant NRF-NRFI2017-01.

# Supplementary Information:

# High Resolution Multispectral Spatial Light Modulators based on Tunable Fabry-Perot Nanocavities

*Shampy Mansha[⊥], Parikshit Moitra[⊥], Xuewu Xu[⊥], Tobias W. W. Mass[⊥], Rasna Maruthiyodan Veetil , Xinan Liang, Shi-Qiang Li, Ramón Paniagua-Domínguez* and Arseniy I. Kuznetsov**

*Institute of Materials Research and Engineering, A*STAR (Agency for Science, Technology and Research), 138634, Singapore*

*Corresponding authors: ramon_paniagua@imre.a-star.edu.sg;*

*arseniy_kuznetsov@imre.a-star.edu.sg*

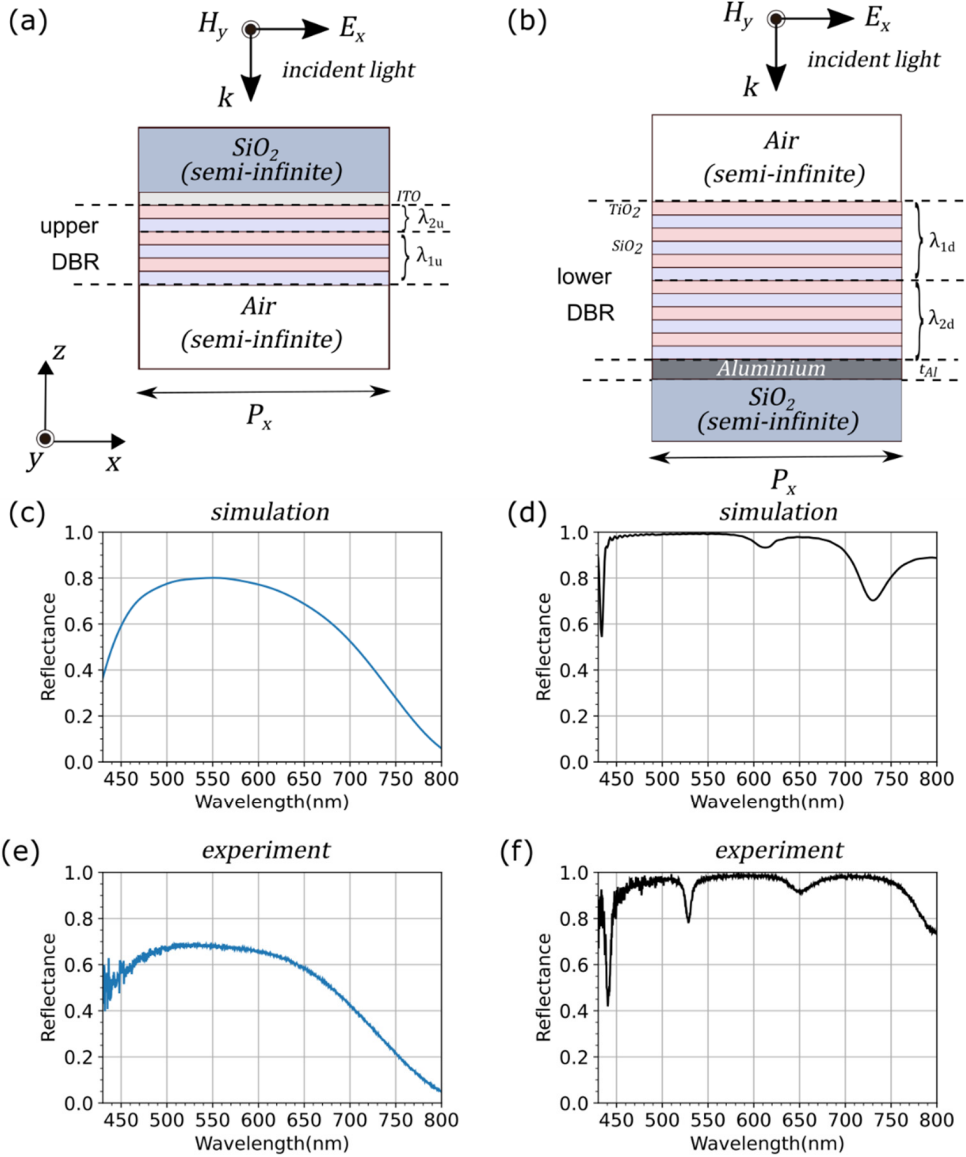

**Figure S1. Reflectance spectra of upper and lower DBRs**: (a) Schematic of the upper DBR, with a semi-infinite layer of glass on the top and a semi-infinite layer of air at the bottom (ITO layer thickness is 23 $nm$). (b) Schematic of the lower DBR, with a semi-infinite layer of air on the top and a semi-infinite layer of glass at the bottom (Al thickness is $t_{Al} = 150\ nm$). In both (a) & (b) there are alternate layers of TiO2/SiO2 ($\lambda/4$: $\lambda_{2u} = 580\ nm$, $\lambda_{1u} = 500\ nm$, $\lambda_{1d} = 450\ nm$, $\lambda_{2d} = 530\ nm$) [$\lambda$ denotes the four target wavelengths for calculating thickness of dielectric layers in DBR stack: where $\lambda_{2u}$ and $\lambda_{1u}$ are for upper DBR stack; and $\lambda_{1d}$ and $\lambda_{2d}$ are for bottom DBR

stack (see Methods: Simulation methodology and optimization of Distributed Bragg Reflector, for more details)] (c), (e) Reflectance spectra of the upper DBR from experiment and simulation, respectively. (d), (f) Reflectance spectra of the lower DBR from experiment and simulation, respectively.

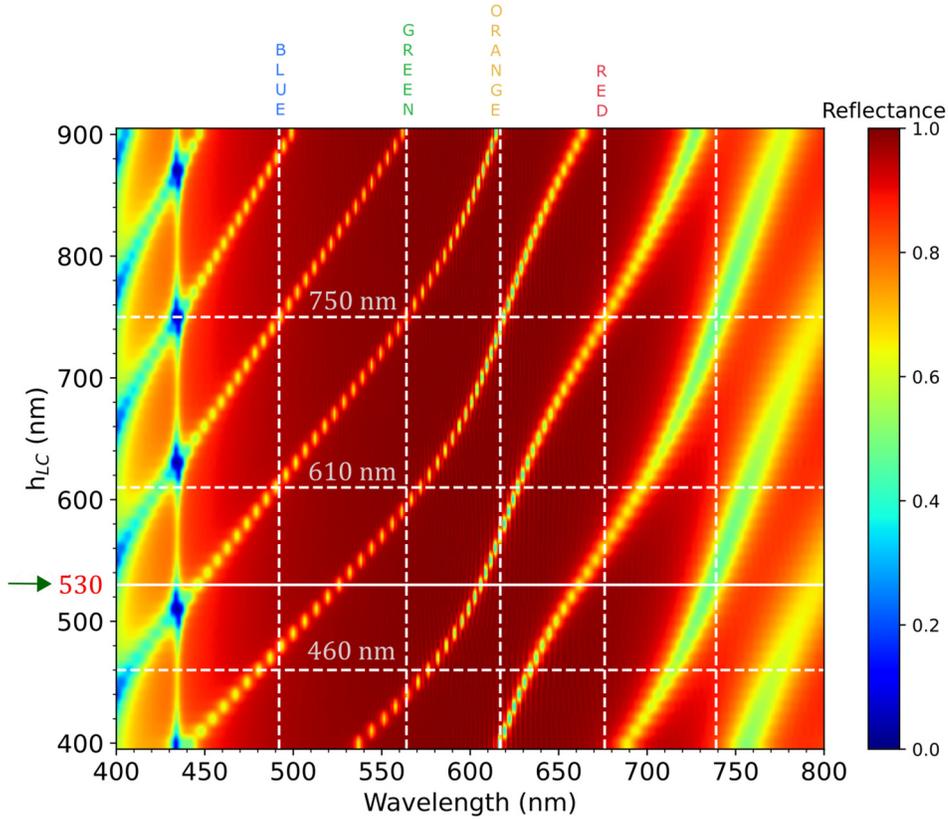

**Figure S2. Reflectance colormap of the FP-SLM obtained from simulations for different heights of the cavity ($h_{LC}$) and wavelengths** (non-pixelated case $P = 1.14\ um$, $gap = 0\ nm$ [where $P$ is the pixel pitch and gap is the inter-pixel gap between adjacent Al electrodes at the bottoms]).

Three different heights can be identified $h_{LC} = 460\ nm,\ 610\ nm,\ 750\ nm$ (dashed white horizontal line) where there is possibility of achieving RGB. The green arrow and the thick white horizontal line indicates $h_{LC} = 530\ nm$, which is the estimated height of the FP-SLM device in the experiment.

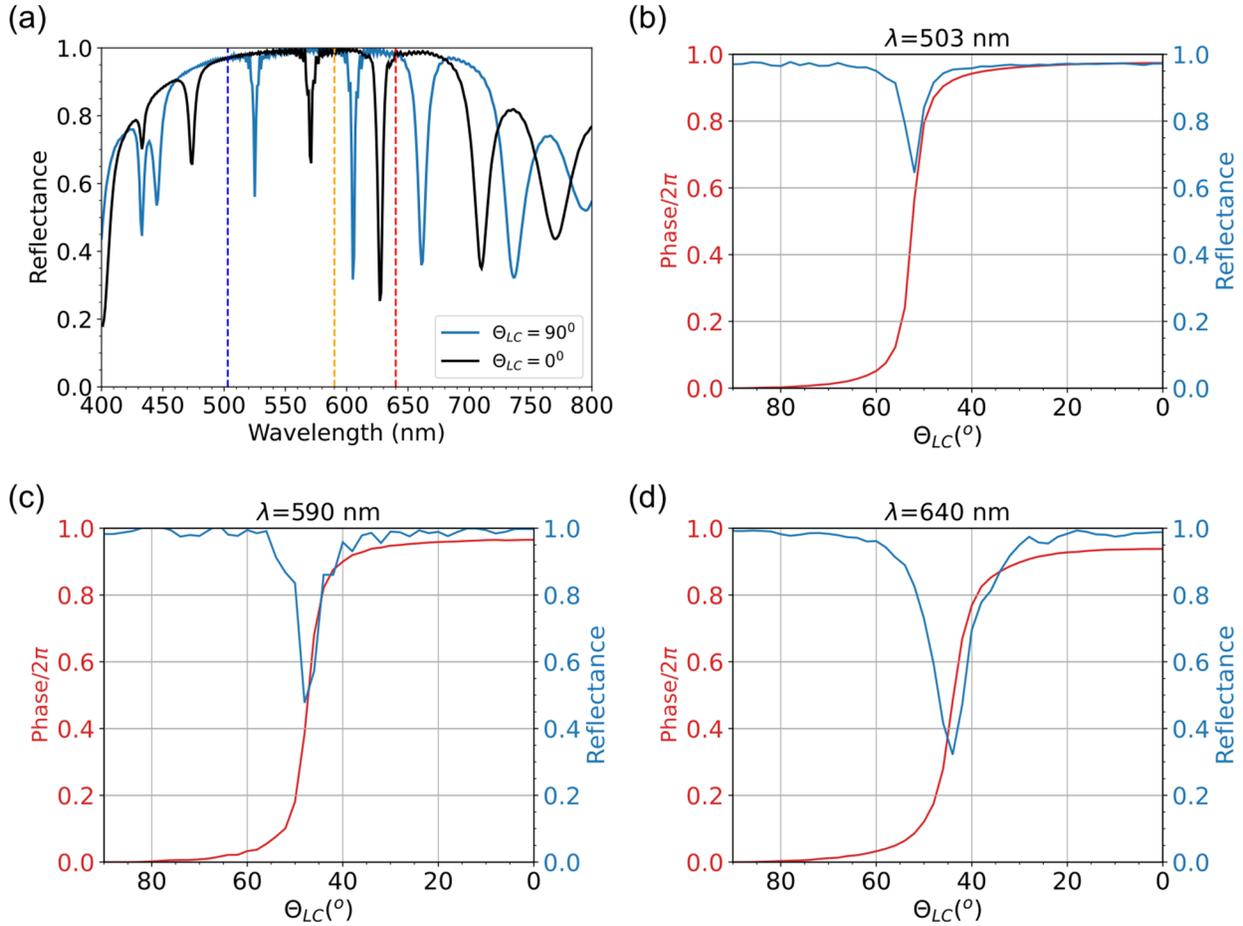

**Figure S3. Reflectance & phase shift in simulation, $h_{LC} = 530\ nm$.** (a) Reflectance spectra for the two cases $\Theta_{LC} = 90^0$ & $\Theta_{LC} = 0^0$. The dashed vertical lines represents the selected wavelengths in blue: 503 nm, orange: 590 nm, red: 640 nm parts of the spectrum. (b)-(d) The reflectance and phase shift plots at the selected wavelengths (as shown in (a)) of 503 nm, 590 nm & 640 nm respectively.

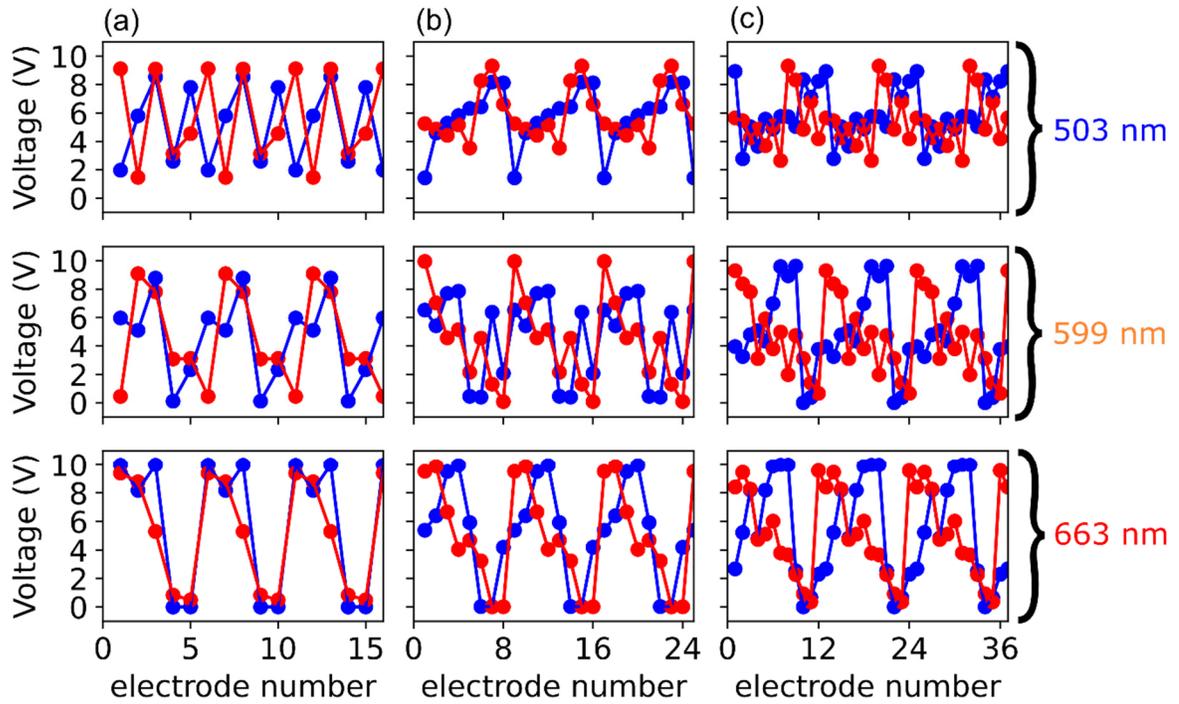

**Figure S4. Optimized voltage profiles for programmable beam steering.** Rows from top to bottom correspond to different wavelengths, as indicated on the right. Columns (a) – (c) show the optimization results for supercell sizes of 5, 8 and 12 electrodes, respectively. Red voltage values correspond to bending into +1 diffraction order and blue voltage values the reverse bending, i.e. into -1 order. This color code corresponds to Fig. 4 in the main text.

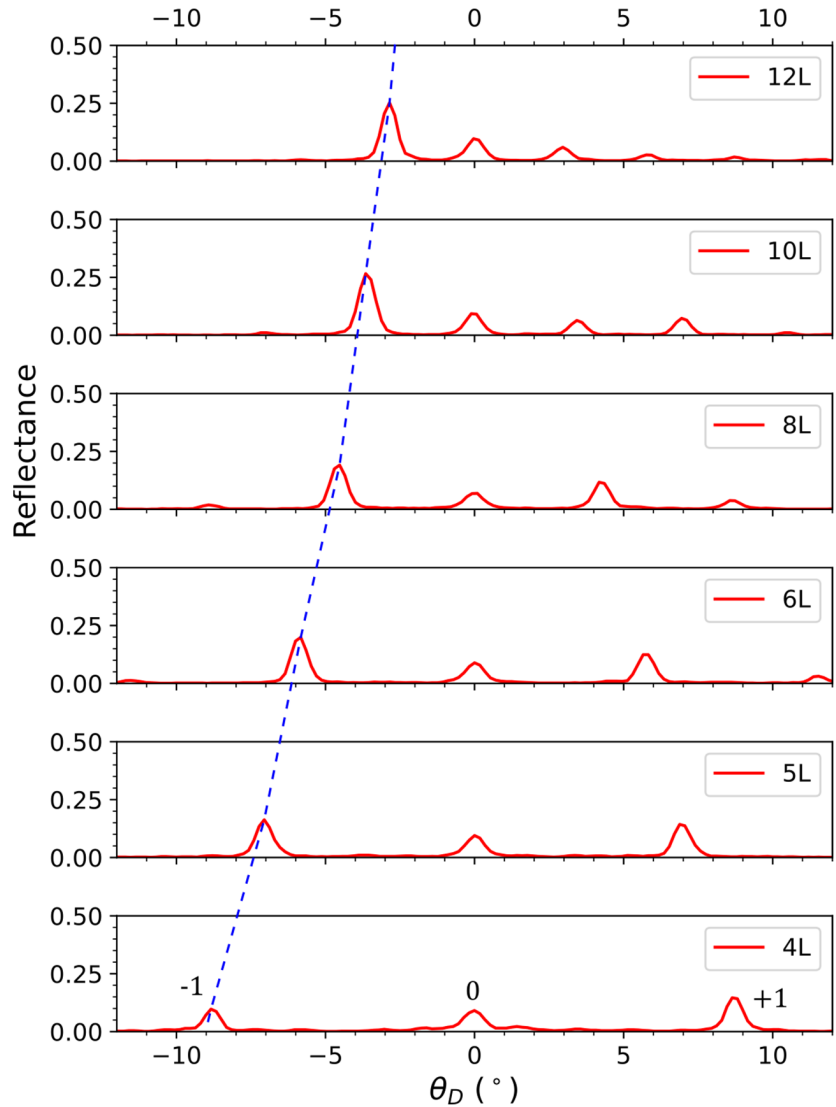

**Figure S5. Programmable beam steering in a wide range of device configurations.** Experimentally measured efficiency as a function of the diffraction angle for the FP-SLM operating at 663 nm. From bottom to top: 4-pixel supercell (4L), 5-pixel supercell (5L), 6-pixel supercell (6L), 8-pixel supercell (8L), 10-pixel supercell (10L) and 12-pixel supercell (12L). The blue dashed line is a guide to the eye tracking the main diffraction order to which power is channeled.

.

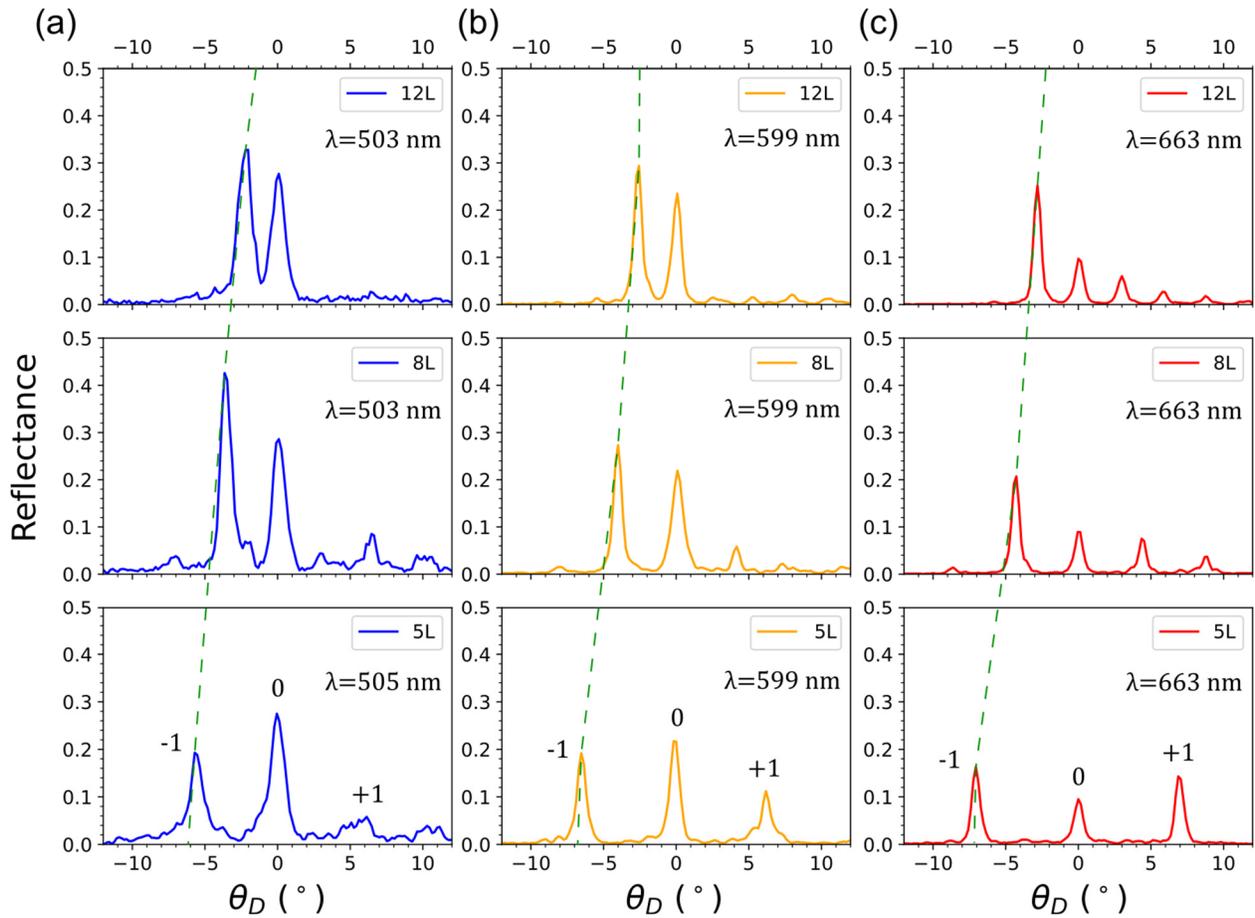

**Figure S6. Multi-spectral programmable beam steering.** Experimentally measured efficiency as a function of the diffraction angle for the FP-SLM operating at wavelengths in the blue (a), orange (b) and red (c) spectral regions. From bottom to top: 5-pixel supercell (5L), 8-pixel supercell (8L) and 12-pixel supercell (12L). The green dashed lines are guides to the eye tracking the main diffraction order to which power is channeled.

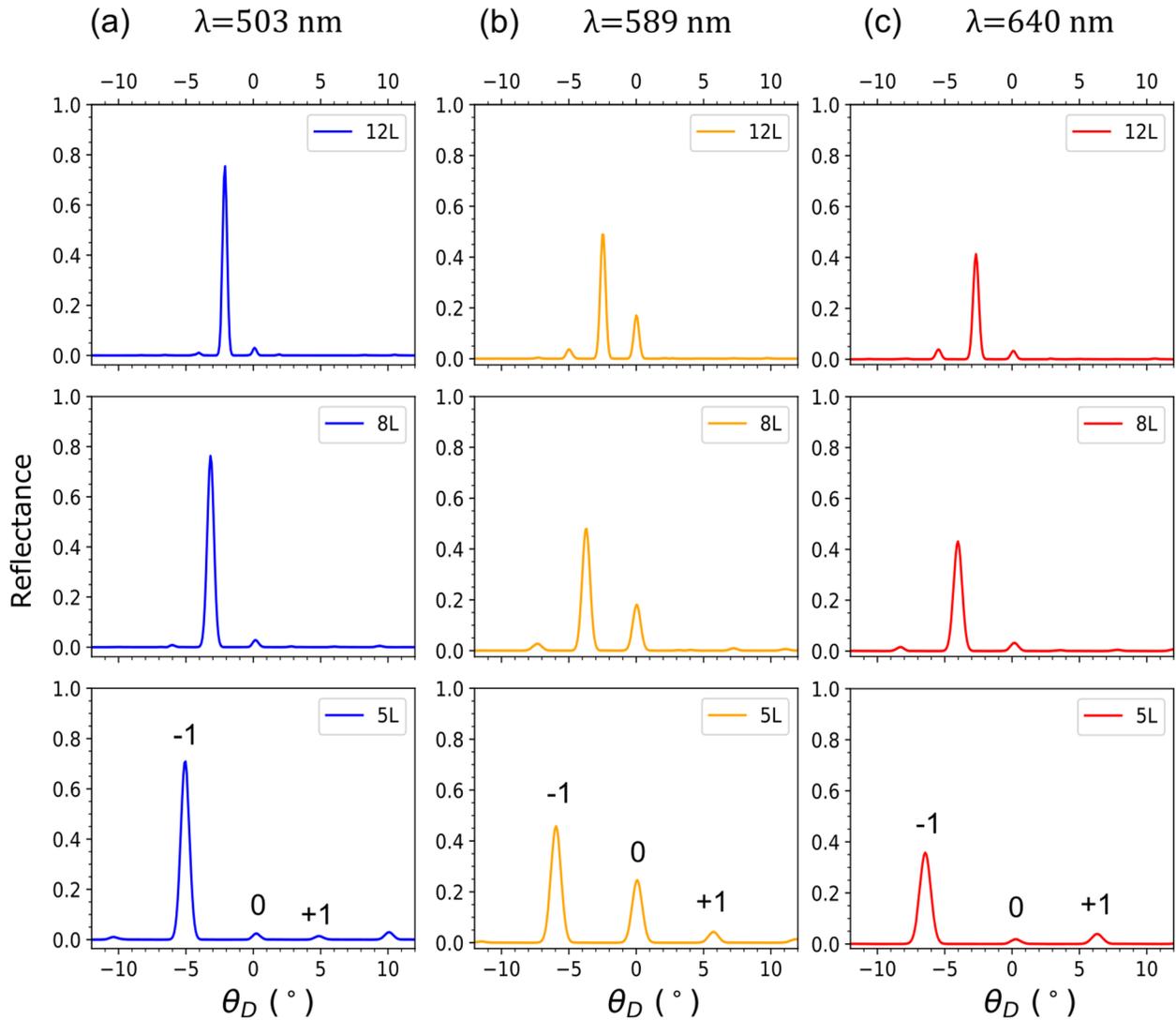

**Figure S7. Simulated multi-spectral programmable beam steering.** Simulated reflectance as a function of the diffraction angle for the FP-SLM operating at wavelengths in the blue (a), orange (b) and red (c) regions. From bottom to top: 5-pixel supercell (5L), 8-pixel supercell (8L) and 12-pixel supercell (12L).

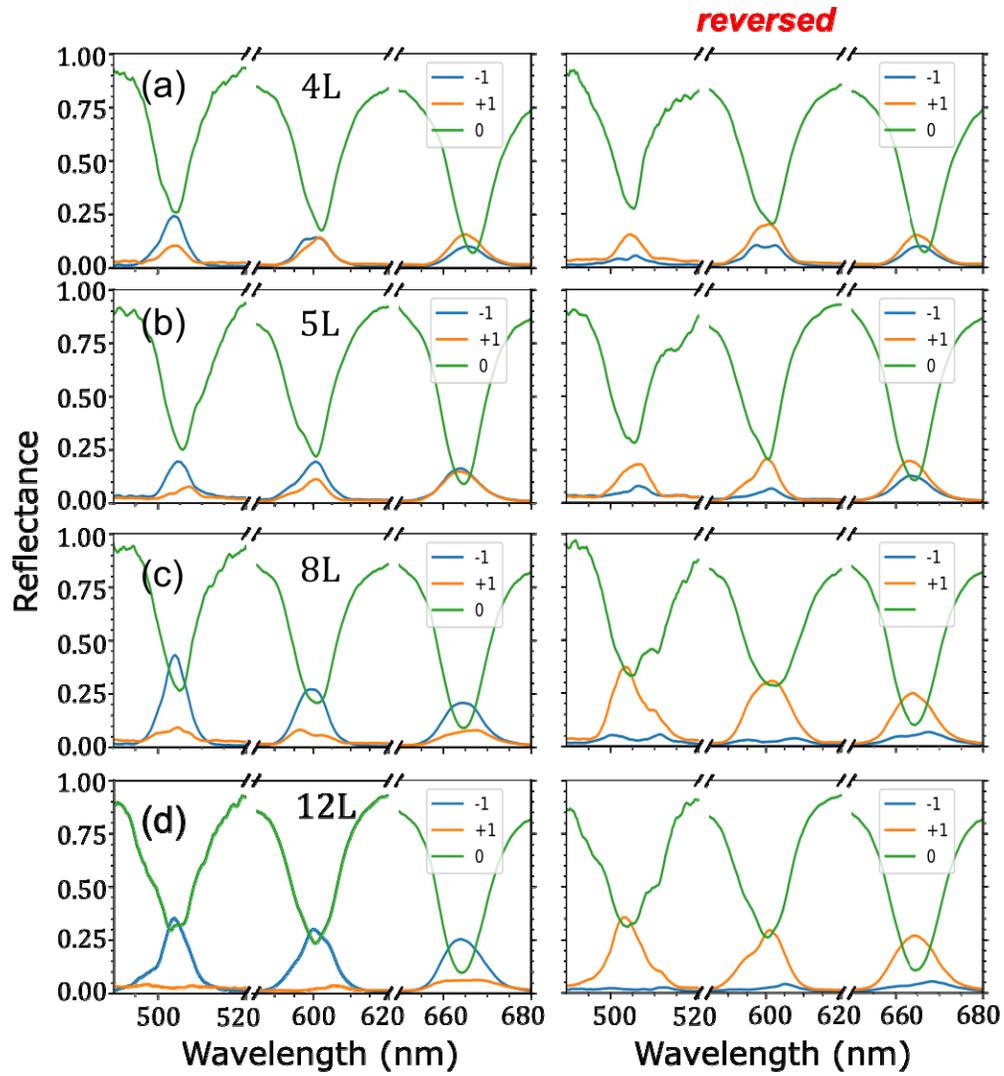

**Figure S8. Experimental efficiencies including 0$^{th}$ order reflection:** (a) 4-pixel supercell, (b) 5-pixel supercell (5L), (c) 8-pixel supercell (8L) and (d) 12-pixel supercell (12L). The left panels correspond to the case in which the device is configured to channel power into the -1$^{st}$ order, while the right ones correspond to the case in which the device is configured to channel power into the +1$^{st}$ order.

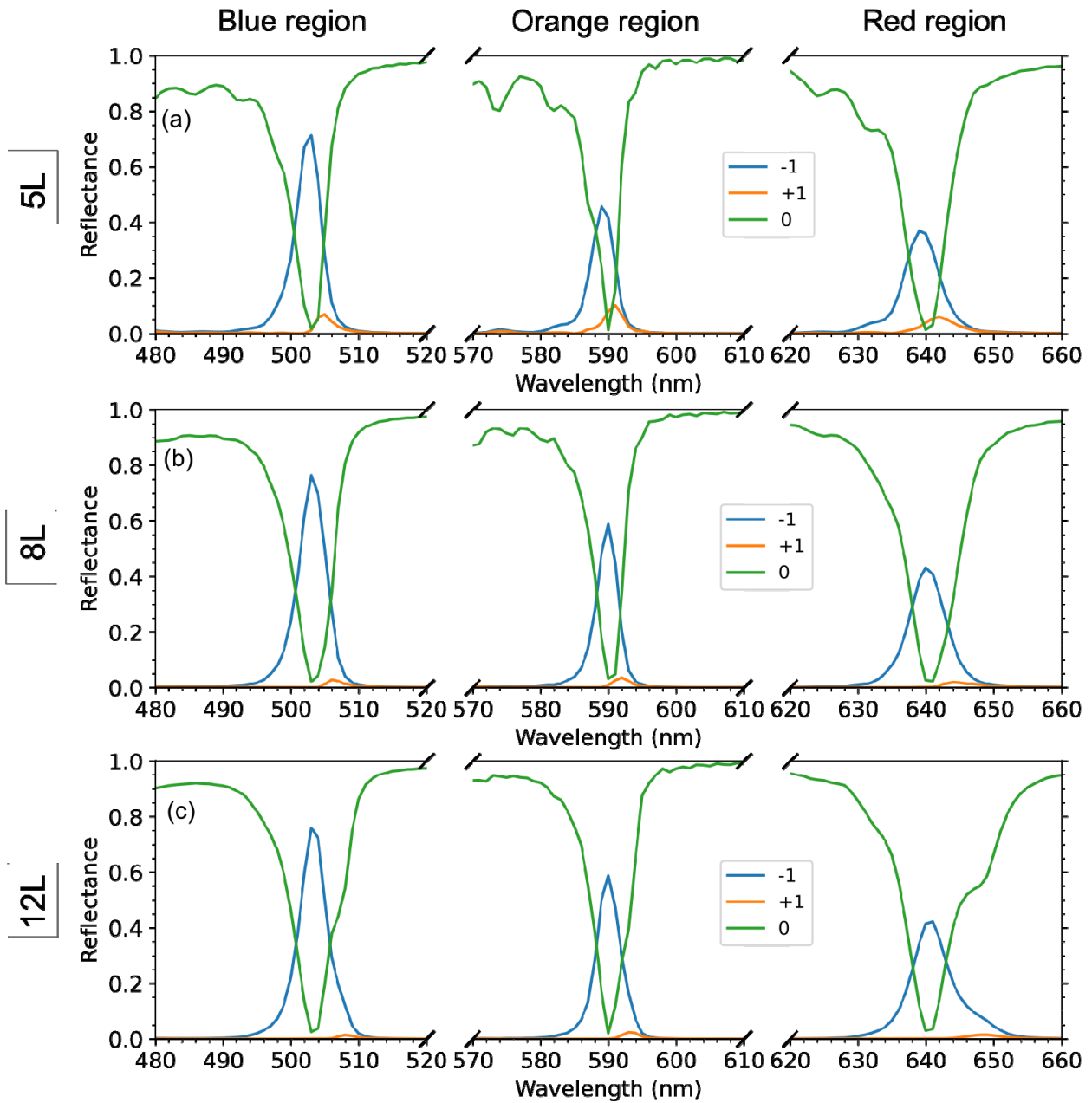

**Figure S9. Simulated efficiencies including 0th order reflection:** (a) 5-pixel supercell (5L), (b) 8-pixel supercell (8L) and (c) 12-pixel supercell (12L). The linear phase (and corresponding LC angles) in simulations are optimized for 503 nm (for blue), 599nm (for orange) and 640 nm (for red).

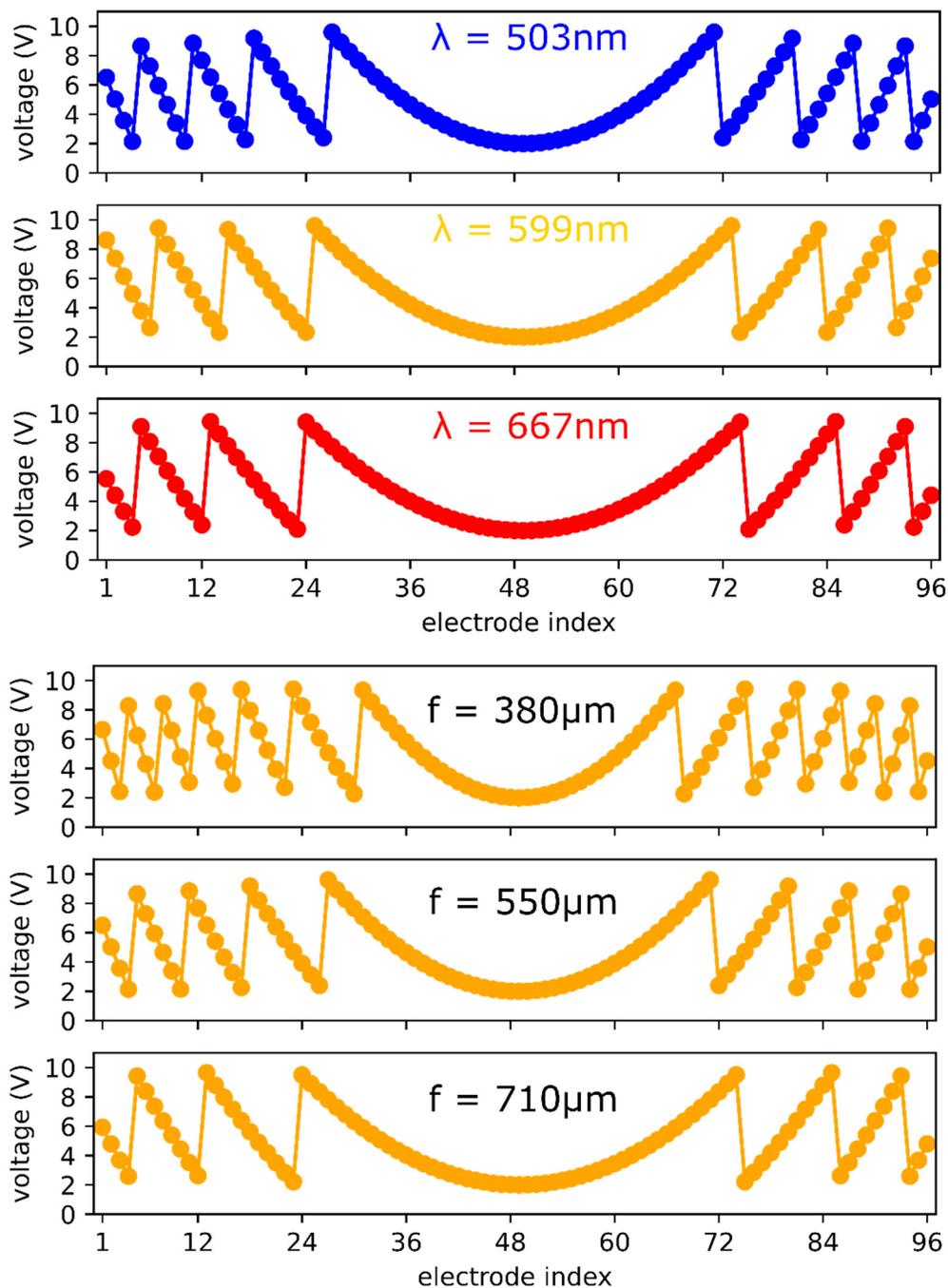

**Figure S10. Voltage profiles used for the lensing experiment.** The first three rows show voltage profiles for lensing at $z = 525$ μm for the three target wavelengths (note that the voltage profiles were obtained for $f = 650\ nm$ but in experiment focusing was obtained at $z = 525\ nm$). These voltage profiles are set to obtain the results presented in Fig. 5(d)-(f) in the main text. The last three rows show lensing profiles for varying focal distance corresponding to Fig. 5 (j)-(l) in the main text. Compared to the target wavelengths and focal lengths, better focusing conditions are obtained for slightly shorter wavelengths and focal distances. We attribute this difference to the fact that the voltages have not been optimized and thus the measured lensing conditions are slightly different.

Captions for video files:

**Supplementary Video S1**: Switching of individual electrodes. The individual electrical switching quality of 96 electrodes in the FP-SLM sample was checked and recorded by applying a voltage of 10 Vrms at 1 kHz to each electrode sequentially under the crossed polarized excitation-collection under a microscope. The LC director was aligned parallel to the grating vector of 1D 96-electrodes and rotated to an angle of 45 degree with respect to both polarizers in the excitation and collection arms. It is shown that almost all the electrodes could be switched individually and the device could be used for beam splitting/bending testing

**Supplementary Video S2**: Spatially resolved reflectance spectra at different distances z from the device. The presented video illustrates the z-scan in Fig. 5 (k) in the main text, for which the voltage pattern with f = 650µm in Fig. S10 has been applied. It can be observed that focusing occurs only around the three operating wavelength ranges. As the voltage pattern is the same for all wavelengths in this measurement, the wavelengths are focused at slightly different z-positions, which can be observed in the video as well. With increasing distances z from the device, first the red wavelengths are being focused, then orange and finally blue.